\documentclass[twocolumn,prl]{revtex4}

\usepackage{epsfig,color}
\usepackage{graphicx}
\usepackage{dcolumn}
\usepackage{bm}
\usepackage{setspace}

\begin{document}
\draft

\title{Fermi liquid behavior of metallic 2D holes at high temperatures}

\author{Xuan P. A. Gao$^{1}$, A. P. Mills Jr.$^{1}$,
A. P. Ramirez$^{2}$, L. N. Pfeiffer$^{3}$, and
K. W. West$^{3}$}
\address{
$^{1}$Physics Department, University of California, Riverside, CA 92507 \\
$^{2}$Los Alamos National Laboratory, Los Alamos, New Mexico 87574\\
$^{3}$Bell Labs, Lucent Technologies, 600 Mountain Avenue, Murray Hill, NJ
07974}

\begin{abstract}
The resistivity $\rho$ of high mobility dilute 2D holes in GaAs
exhibits a peak at a certain temperature $T^*$ in zero magnetic
field($B$=0). In the $T>T^*$ regime where d$\rho$/d$T<$0, we
observe for the first time both the $\nu=1$ quantum Hall(QH)
effect and a low field insulator-QH transition which is consistent
with the 2D hole system being a Fermi liquid(FL). The known linear
$T$-dependent conductivity in this $T$ regime then can be
explained by hole-hole Coulomb interactions of this FL. The fact
that the system is metallic (d$\rho$/d$T>$0) for $T<T^*$ implies
that the high temperature FL transforms into the 2D metallic state
in the neighborhood of $T^*$.

\end{abstract}
\maketitle

The effect of disorder on the electronic properties of solids has
been one of the central problems in modern condensed matter
physics\cite{Lee&rama}. Anderson's pioneering work demonstrated
the localization effect of disorder on the electron
wavefunctions\cite{andersonlocalization}. The seminal
one-parameter scaling theory of localization by Abrahams {\it et
al.} made the remarkable conclusion that all non-interacting
disordered electronic systems are localized in zero magnetic field
in two dimensions(2D)\cite{scaling79}. However, when a strong
perpendicular magnetic field is applied to 2D systems, quantized
Landau Levels(LLs) form and extended states can exist around the
centers of the LLs. In high quality 2D systems, the integer QH
effect can be seen  at low temperatures when the Fermi energy
moves across the mobility edges which separate the extended states
in the centers of LLs from the localized states between
LLs\cite{QH,LaughlinQH}.

To connect the $B$=0 insulating ground state in 2D and the QH
states  in finite $B$, it was argued that the positions of the
extended states should float up in energy as
$B\to$0\cite{floatup}. Therefore a zero field insulator-QH
transition should be seen when an extended state moves across the
Fermi energy by increasing $B$ from zero. The $B$=0 Anderson
insulator to QH transition was first demonstrated in a $B$=0
strongly localized 2D electron system\cite{Jiang93} and later in
$B$=0 weakly localized 2D systems as well\cite{floatupweak}.
Moreover, as predicted by the theory, experiment found the
positions of the extended states deviating from the LLs and
floating up above the Fermi surface as $B\to$0 in a 2D electron system
with tunable density\cite{Glozman}.

The recent discovery of the possible $B$=0 2D metallic state and
metal-insulator transition(MIT) in low density 2D electron/hole
systems has attracted a lot of attention\cite{mitreview}. Note
that $r_s$, the dimensionless ratio between the Coulomb
interaction energy and the Fermi energy of these dilute systems
showing a 2D MIT, is much larger than one. This poses the question
if the the conventional theory of localization is applicable and
what is the role of strong Coulomb interactions  in these dilute
2D systems. At present, there is still no consensus on either the
origin of the metallic behavior or the nature of the insulating
state of the $B$=0 2D MIT. Nonetheless, Shubnikov-de Haas(SdH)
oscillations have been seen on both sides of the MIT and the
insulating state of the MIT can be turned into a QH conductor by
applying a perpendicular magnetic field, that is quite similar to
conventional Fermi liquid type insulating 2D systems. The
insulator-QH transition in a perpendicular magnetic field was
employed on the insulating side of the 2D MIT by different authors
to investigate how the extended states behave in the
density-magnetic field phase
diagram\cite{Haneinnature,Dultz,Yasin}.

In this paper, we report for the first time the observation of the
insulator-QH transition at high temperatures in a dilute 2D hole
system that is metallic at low $T$.  All previous studies on the
insulator-QH transition were done on true $B$=0 2D insulators,
either in the conventional Fermi liquid\cite{Jiang93,floatupweak}
or the insulating phase of the 2D
MIT\cite{Haneinnature,Dultz,Yasin}. This paper focuses on the
metallic side of the 2D MIT, where the zero field $T$-dependent
resistivity $\rho(T)$ changes from metallic to insulating-like
above a certain temperature scale
$T^*$\cite{HaneinholePRL,MillsPRL99,Lilly,Noh03,Gaothesis}. The
conductivity of 2D holes was shown to be a linearly increasing
function of $T$ in the $T>T^*$ insulating
regime\cite{Noh03,Gaothesis}. There exist a few theoretical models
to explain the insulating $\rho(T)$ of the 2D metallic state at
$T>T^*$. A simple non-interacting picture for the apparent
insulating behavior of metallic 2D systems at high $T$ is the
$T$-dependent scattering of a non-degenerate system by Das Sarma
and Hwang\cite{DasSarma}. On the other hand, Spivak conjectured
that Coulomb correlations could be responsible for such an
insulating $\rho(T)$\cite{Spivak}.

\begin{figure}[btph]
\centerline{\psfig{file=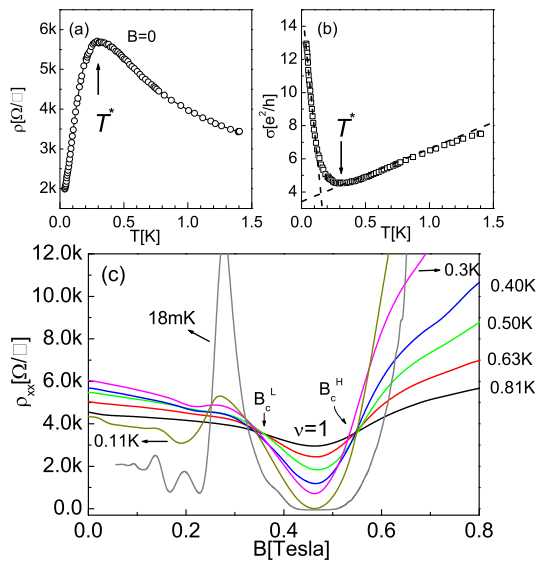,width=8cm}}
 \caption{ (a)Temperature dependent resistivity $\rho(T)$ of 2D
holes in a 10nm wide GaAs quantum well with density
$p$=1.13$\times$10$^{10}$cm$^{-2}$ in zero magnetic field.  Note
that $\rho(T)$ changes from insulating-like to metallic below
$T^*\approx$0.3K. (b) The same data as in (a) plotted as
conductivity in units of $e^2/h$ vs. $T$. (c)Longitudinal
magneto-resistivity $\rho_{xx}(B)$ of 2D holes in (a) and (b) at
various temperatures. For temperatures above $T^*$, the
$\rho_{xx}(B)$ curves show a low $B$ insulator-$\nu$=1 QH
transition at $B_c^L$=0.35T. For comparison, we also show
$\rho_{xx}(B)$ curves at 0.11K and 18mK, which are below $T^*$. }
  \label{fig1}
\end{figure}

Fig.\ref{fig1}c, to be discussed in detail below, demonstrates the
existence of SdH oscillations as well as an insulator-QH
transition in the high $T$ insulating regime of metallic 2D holes.
The existence of a quantum magneto-resistance minimum and an
insulator-QH transition suggest the metallic 2D hole system in
fact behaves like a normal insulating Fermi liquid at $T>T^*$.
Together with the known linear $T$-dependent zero field
conductivity[Fig.\ref{fig1}a,b], our results suggest the
insulating $\sigma(T)$ at $T>T^*$ is due to the Coulomb
interactions of Fermi liquid in the "ballistic" transport
regime\cite{Zala}. Therefore, the metallicity of the system for
$T<T^*$ implies that the 2D metallic state is fundamentally
different from the high temperature insulating Fermi liquid.

Our measurements were performed on a back-gated 2D hole system in
a 10nm  wide GaAs quantum well made from one of the wafers used in
our previous study \cite{GaoPRL03}, a (311)A GaAs wafer using
Al$_{.1}$Ga$_{.9}$As barriers and symmetrically placed Si
delta-doping layers. The wafer was thinned to $ \approx $150 $\mu
$m  for gating from the back of the sample and prepared in the
form of a Hall bar, of approximate dimensions (2.5$ \times $9)
mm$^{2}$, with diffused In(5\%Zn) contacts. The measurement
current was applied along the [\={2}33] direction and kept low
such that the power delivered on the sample is less than a few
fWatts/cm$^2$ to avoid over heating the holes.

Fig.\ref{fig1}a presents the zero magnetic field temperature
dependent resistivity of 2D holes with density
$p$=1.13$\times$10$^{10}$cm$^{-2}$  which is on the metallic side
of the MIT. For this sample, the MIT happens around $p_c$=0.78
$\times$10$^{10}$cm$^{-2}$ when the density is changed by the
backgate. It can be seen in Fig.\ref{fig1}a that $\rho(T)$ changes
from insulating-like(d$\rho$/d$T<$0) to metallic(d$\rho$/d$T>$0)
below $T^*\approx0.3K$. In Fig.\ref{fig1}b we plot the data as
conductivity $\sigma(T)$, in units of $e^2/h$. For either
$T>1.5T^*$ or $T<0.5T^*$, $\sigma(T)$ can be viewed as a linear
function of $T$, but with a different coefficient
d$\sigma(T)$/d$T$ \cite{Noh03,Gaothesis}. Fig.\ref{fig1}c shows
the main result of this paper, the insulator-QH transition in the
high temperature insulating regime of the metallic state. The
longitudinal magneto-resistivity $\rho_{xx}(B)$ at several
temperatures are plotted on Fig.\ref{fig1}c. For $T>T^*$, there is
a low field insulator-$\nu=1$ QH transition at critical field
$B_c^L$=0.35Tesla, denoted by the crossing of $\rho_{xx}(B)$
curves. The $\rho_{xx}(B)$ curves also cross at a high critical
field $B_c^H$=0.55Tesla. For comparison, we also include
$\rho_{xx}(B)$ curves at 0.11K and 18mK which are below $T^*$. As
$T$ drops below $T^*$ the size of the low field negative
magneto-resistivity decreases and a sharp resistivity spike
develops around $\nu=5/3$. The large resistivity spike at
$\nu\approx 5/3$ and how the $B$=0 2D metal transforms into QH
states at low $T$ need further study. Again, we emphasize that
although SdH oscillations are known to exist in the 2D metallic
state at low $T$, to the best of our knowledge Fig.\ref{fig1} is
the first report of a SdH minimum and an insulator-QH transition
in the high $T$($>T^*$) insulating regime of the 2D holes with
$p>p_c$.

Although the $\rho_{xx}(B)$ data and insulator-QH transition in
Fig.\ref{fig1} appear to be similar to previously studied Anderson
insulator-QH transitions \cite{Jiang93,floatupweak}, the
underlying physics should be very different. Most of the previous
experiments were performed in the low temperature diffusive regime
such that $k_BT<\hbar/\tau$, where $k_B,\hbar,\tau$ are the
Boltzmann constant, Planck's constant divided by 2$\pi$ and the
scattering time, respectively. Thus the low $B$ insulating
$\rho(T)$ was likely to be mainly driven by single particle
quantum interference (Anderson localization). In the temperature
regime($T>T^*$) where our high mobility metallic sample shows the
"insulator-QH transition", there is no single particle
interference effect due to strong dephasing, and the sample is
actually in the ballistic regime ($k_BT>\hbar/\tau$)\cite{Zala}.
Recent interaction theories on the conductivity of 2D Fermi
liquids in the ballistic regime predict a linear $T$-dependent
zero field conductivity\cite{Zala} and a parabolic negative
magneto-resistivity at $\omega_c\tau>$1\cite{Gornyi}. The data in
Fig.\ref{fig1} show that our sample behavior is qualitatively
consistent with the predictions of ref.\cite{Zala,Gornyi} at
$T>T^*$\cite{Li}. Recently, some authors have attempted to
interpret the $T<T^*$ metallic behavior using the Fermi liquid
theory in ref.\cite{Zala}. Under this interpretation, we found the
fitted Fermi liquid parameters to be inconsistent between 2D hole
samples with similar densities but different
mobilities\cite{Gaothesis}. Note that since d$\sigma(T)$/d$T$
changes sign at $T^*$, it is not a priori known whether the
$\sigma(T)\sim T$ at $T>T^*$ or at $T<T^*$ is due to interaction
corrections. We believe that the $\sigma(T)\sim T$ at $T>T^*$ is
related to interaction effects of a ballistic Fermi liquid, where
both the parallel field and perpendicular field effect are
consistent with theoretical expectations\cite{note}. {\narrowtext
\begin{figure}[htb]
\epsfxsize=7cm \centerline{\epsfbox{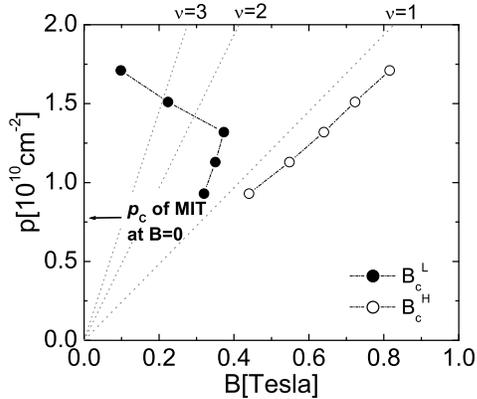}} \caption{The
lower critical field $B_c^L$ and higher critical field $B_c^H$  of
the high $T$ insulator-QH-insulator transition for 2D metallic
holes with various densities. Black lines connecting data points
are only guides to the eye. As a reference, we show the positions
of SdH dips for $\nu=1,2,3$ as dotted lines on the
density-magnetic field diagram.}\label{fig2}
\end{figure}
}

Observing the SdH oscillation and an "insulator-QH transition" in
the $T>T^*\sim T_F$ insulating regime of low density 2D holes is
somewhat unexpected, since it was originally believed that the
insulating $\rho(T)$ at $B=0$ in this temperature range is related
to the non-degenerate nature of the low density hole
gas\cite{DasSarma}. Note that if the insulating $\rho(T)$ at
$T>T^*$ has  a classical origin as in ref.\cite{DasSarma}, it
would be unlikely for the sample to show a negative
magneto-resistivity and the "insulator-QH transition" in
Fig.\ref{fig1} because the classical Drude magneto-resistivity is
zero. At present, it is unclear if a more elaborated
semi-classical model can reproduce a non-monotonic $T$ dependent
zero field $\rho(T)$ and an "insulator-QH transition" at high $T$
at the same time. To further elucidate the high $T$ "insulator-QH"
transition of metallic 2D holes, we plot both $B_c^L$ and $B_c^H$
for five densities on the density-magnetic field diagram shown in
Fig.\ref{fig2}. The values of $B_c^L$ and $B_c^H$ are obtained
from the crossing of the $\rho_{xx}(B)$ curves at temperatures
$T>T^*$. For $p$=0.93,1.13,1.32$\times 10^{10}$cm$^{-2}$, the
insulator-$\nu=1$ QH transition is observed at $B_c^L$, while for
higher densities the sample enters the QH state with $\nu>1$ at
$B_c^L$. As a reminder, the density dependence of $B_c^L,B_c^H$ in
Fig.\ref{fig2} is related to the nature of the high $T$ insulating
phase of 2D holes on the metallic side($p>p_c$) of the MIT,
differing from all previous insulator-QH studies which were on the
insulating($p<p_c$) side of the $B=0$
MIT\cite{Haneinnature,Dultz,Yasin}. As we suggest above, the zero
field localization exhibited by the sample at $T>T^*$ could be due
to the Coulomb interactions of a "ballistic" Fermi liquid. Thus,
the trace of $B_c^L, B_c^H$ in Fig.\ref{fig2} represents a
metal-insulator boundary in $p,B$ space for a Fermi liquid in
which localization is prompted merely by interactions. Zhang {\it
et al.} has constructed a global phase diagram for a 2D system in
a magnetic field\cite{Zhang}, considering only the disorder
induced Anderson localization. An extended theoretical phase
diagram like ref.\cite{Zhang} including the interaction effect has
yet to be developed to make comparison with Fig.\ref{fig2}.

The persistence of the $\nu=1$ QH effect into the $T\sim T_F$
regime  suggests a large energy gap $\Delta_1$ of the $\nu=1$ QH
state for the low density 2D hole gas. We plot $\rho_{xx}(T)$ at
the center of the $\nu=1$ QH state for 2D holes with different
densities on Fig.\ref{fig3}. The black lines are fitted curves
using the activated function $exp(-\Delta_1/2T)$. The density
dependence of $\Delta_1$ is presented in the inset of
Fig.\ref{fig3}. The Fermi temperature $T_F$ is also plotted using
effective hole masses $m^*=0.38m_e$ and $0.19m_e$ to compare with $\Delta_1$.
Fig.\ref{fig3} shows that the energy gap in the $\nu=1$ QH state could
be 2-3 times larger than the Fermi energy of 2D holes in the
low density regime. Many body effects have been known to be
important in the $\nu=1$ QH state where novel collective spin
excitations named Skyrmions may exist\cite{Sondhi,Schmeller}.
Therefore $\Delta_1$ should be related to the interaction energy,
which is much larger than the Fermi energy in low density holes.
Interestingly, the inset of Fig.\ref{fig3} shows a sharp
non-monotonic density dependence of $\Delta_1$. We suspect the
sharp drop of $\Delta_1$ at lower density is due to the
competition between the QH liquid and Wigner solid at the lowest
Landau level\cite{Shayegan}.
{\narrowtext
\begin{figure}[htb]
\epsfxsize=7cm \centerline{\epsfbox{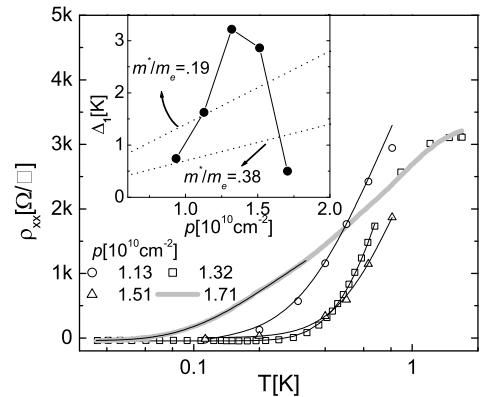}}
\caption{$\rho_{xx}(T)$ of metallic 2D holes in 10nm wide GaAs
quantum well  at the center of the $\nu=1$ QH state. Data for four
different densities are presented. The black lines are fitted
curves using the thermally activated form $exp(-\Delta_1/2T)$. The
inset presents the density dependence of $\Delta_1$, the energy
gap of $\nu=1$ QH state. The dotted lines in the inset show the
Fermi temperatures, using hole mass $m^*=0.19m_e$ or $0.38m_e$.
}\label{fig3}
\end{figure}
}

Finally we would like to comment on the contradictory results
concerning the fate of the lowest extended states as $B\to0$ in
low density GaAs 2D electron/hole systems in
ref.\cite{Haneinnature,Dultz,Yasin}. Both Hanein {\it et al.}
\cite{Haneinnature} and Yasin {\it et al.} \cite{Yasin} used the
low $B$ insulator-QH transition to track the positions of the in
2D GaAs systems in the insulating side of MIT. The low $B$
insulator-QH transition in the $T>T^*$ regime observed in this
paper suggests that caution must be taken while interpreting the
finite temperature insulator-QH transition. Due to the
non-monotonicity of $\rho(T)$ in dilute 2D GaAs systems, the
critical density/resistivity of the MIT is not well defined and
could depend on the experimental temperature. If an experiment is
done at high temperature, the critical density could appear to
have a higher value  with a correspondingly lower value of
critical resistivity. As a result, for experiments where
temperature is high or the electrons/holes are overheated, the
insulator-QH transition for $p$ slightly lower than $p_c$ could be
only the insulator-QH transition of {\it metallic} holes at
$T>T^*$. In fact, the $B_c^L$ of the high $T$ insulator-QH
transition increases with $p$ for $p$ just above $p_c$, as
Fig.\ref{fig2} shows, which may cause the $B_c^L$ appear to "float
up" for $p<p_c$ in a sample with high $T$. Thus, the positions of
extended states extracted from the insulator-QH transition depend
critically on the sample temperature around the region $p\sim
p_c$, which is exactly the crucial density region where the
extended states appear to float up\cite{Yasin} or
saturate\cite{Haneinnature,Dultz}. We note that $T^*$ for 2D holes
in GaAs was found to be linearly dependent on density and to
extrapolate to zero at some density $p^*$\cite{GaoPRL03}.
Practically the positions of extended states determined from
insulator-QH transitions can only be reliably extrapolated to
$T\to 0$ for densities below $p^*$ but not the $p_c$ determined
from the temperature coefficient of zero field $\rho_{xx}(T)$. In
our opinion, the different behavior of the lowest extended states
as $B\to 0$ around the $p_c$ of the 2D MIT in
ref.\cite{Haneinnature,Dultz,Yasin} might be simply related to the
different base temperatures of the 2D systems.

In summary, we demonstrate for first time that there exist SdH
oscillations and an insulator-QH transition in the high
temperature insulating regime of low density (low $T$-metallic) 2D
holes in a GaAs quantum well. We also show that the energy gap of
the $\nu=1$ QH state has a large value, attesting the strong
interactions in the low density GaAs hole system. These results
together with the linear $T$ dependent conductivity suggest the 2D
metallic hole system behaves like a localized Fermi liquid in the
ballistic transport regime for $T$ higher than $T^*$, the
temperature at which zero field resistivity changes from
insulating to metallic.

The work at UCR is supported by a LANL CARE program.


\begin{references}
\bibitem{note} Fitting $\sigma(T)$ of 2D metallic state at $T>T^*$ into the theory in ref.\cite{Zala} yields a small negative Fermi liquid parameter $F_0^\sigma$, which decreases towards zero as the critical density of MIT is approached. The parallel magnetic field effect at $T>T^*$ also agrees with the $F_0^\sigma$ fitted from zero field $\sigma(T)$. See discussions in ref.\cite{Gaothesis}.
\end{references}
\end{document}